\documentclass[prd,preprint,superscriptaddress]{revtex4-1}
\usepackage{graphicx}
\usepackage{latexsym}
\usepackage{amsmath}
\usepackage{amssymb}

\begin{document}
\title{Cosmology in massive gravity}

\author{Yungui Gong}
\affiliation{MOE Key Laboratory of Fundamental Quantities Measurement, School of Physics, Huazhong University of Science and Technology, Wuhan 430074, China}
\affiliation{Institute of Theoretical Physics, Chinese Academy of Sciences, Beijing 100190, China}
\email{yggong@mail.hust.edu.cn}

\begin{abstract}
We argue that more cosmological solutions in massive gravity can be obtained if the metric tensor and the tensor $\Sigma_{\mu\nu}$ defined
by St\"{u}ckelberg fields take the homogeneous and isotropic form. The standard cosmology with matter and radiation dominations in the
past can be recovered and $\Lambda$CDM model is easily obtained. The dynamical evolution of the universe is modified at very early times.
\end{abstract}


\preprint{arXiv: 1207.2726}
\date{\today}

\maketitle

\section{Introduction}

The discovery of accelerating expansion of the universe by
the observations of Type Ia supernovae in 1998 \cite{acc1,acc2} motivated the search for dark energy
and modified gravity. The gravitational force decays at a scale larger than $m^{-1}$ if graviton has
a mass $m$, so massive gravity may be used to explain the cosmic acceleration. Naively, the mass of graviton
should be very small so that gravity is still approximately a long range force, therefore,
it is expected that the mass of graviton is about Hubble scale $m\sim H_0$.
Dvali, Gabadadze and Porrati proposed that general relativity is modified
at the cosmological scale \cite{dgp}. In this model, there are a continuous tower of massive
gravitons. The first attempt of a theory of gravity with massive graviton
was made by Fierz and Pauli \cite{pauli}. However, the linear theory with the Fierz-Pauli mass is in
contradiction with solar system tests \cite{vdvz1,vdvz2}.  Recently, de Rham, Gabadadze and Tolley
introduced a nonlinear theory of massive gravity \cite{massgrav} that is free from Bouldware-Deser ghost \cite{bdghost,hassan}.
The cosmological solutions for massive gravity were sought in \cite{amico,koyama1,koyama2,gumruk11,gratia12,kobayashi12,volkov12,mukohyama,tolley,langlois12,vonStrauss:2011mq,Berezhiani:2011mt,Akrami:2012vf,Koyama:2011wx,Motohashi:2012jd,Tasinato:2012ze}.
The first homogenous and isotropic solution was found for spatially open universe in \cite{gumruk11} and the massive graviton
term is equivalent to a cosmological constant. The same solutions were then found for spatially open and closed universe
in \cite{gratia12,kobayashi12}. In addition to the equivalent cosmological constant solution, more general cosmological solutions were also found in \cite{tolley,langlois12}
by taking the de Sitter metric as the reference metric.
We follow the approach in \cite{tolley,langlois12} and proposed a new approach to find more general cosmological solutions.

\section{massive gravity}

The theory of massive gravity is base on the following action \cite{massgrav}
\begin{equation}
\label{action}
S=\frac{M_{pl}^2}{2}\int d^4x\sqrt{-g}(R+m_g^2\mathcal{U})+S_m,
\end{equation}
where $m_g$ is the mass of the graviton, the mass term
\begin{equation}
\label{massv}
\mathcal{U}=\mathcal{U}_2+\alpha_3\mathcal{U}_3+\alpha_4\mathcal{U}_4,
\end{equation}
\begin{gather}
\label{massv2}
\mathcal{U}_2=[\mathcal{K}]^2-[\mathcal{K}^2],\\
\label{massv3}
\mathcal{U}_3=[\mathcal{K}]^3-3[\mathcal{K}][\mathcal{K}^2]+2[\mathcal{K}^3],\\
\label{massv4}
\mathcal{U}_4=[\mathcal{K}]^4-6[\mathcal{K}]^2[\mathcal{K}^2]+8[\mathcal{K}^3][\mathcal{K}]-6[\mathcal{K}^4],
\end{gather}
and
\begin{equation}
\label{tensor1}
\mathcal{K}^\mu_\nu=\delta^\mu_\nu-(\sqrt{\Sigma})^\mu_\nu,
\end{equation}
The tensor $\Sigma_{\mu\nu}$ is defined by four St\"{u}ckelberg fields $\phi^a$ as
\begin{equation}
\label{stuckphi}
\Sigma_{\mu\nu}=\partial_\mu\phi^a\partial_\nu\phi^b\eta_{ab}.
\end{equation}
The reference metric $\eta_{ab}$ is usually taken as the Minkowski one.
The cosmological solution was first found in \cite{gumruk11} for an open universe and the mass term
behaves like an effective cosmological constant with
\begin{equation}
\label{efflambda}
\begin{split}
\Lambda_{eff}=&-m^2_g\left(1+3\alpha_3\pm\sqrt{1+3\alpha_3+9\alpha_3^2-12\alpha_4}\right)\times\\
&\frac{(1+9\alpha^2_3-24\alpha_4\pm(1+3\alpha_3)\sqrt{1+3\alpha_3+9\alpha_3^2-12\alpha_4})}{9(\alpha_3+4\alpha_4)^2}.
\end{split}
\end{equation}
The same solution was then found in \citep{gratia12} for a flat universe
by considering an arbitrary spatially isotropic metric and a spherically symmetric ansatz for the St\"{u}ckelberg fields.
In \cite{Motohashi:2012jd}, the authors obtained the solution by assuming isotropic forms for both the physical and reference metrics.
For a general case with positive, negative and
zero curvature, Kobayashi {\it et al.} found the same solution with $\Lambda_{eff}=m^2_g/\alpha$ for the particular choices
of parameters $\alpha_3$ and $\alpha_4$ \cite{kobayashi12},
\begin{equation}
\label{1parm}
\alpha_3=\frac{1}{3}(\alpha-1),\quad \alpha_4=\frac{1}{12}(\alpha^2-\alpha+1).
\end{equation}

In \cite{tolley,langlois12}, the authors assumed that the spacetime metric takes the form
\begin{equation}
\label{frwmetric}
ds^2=g_{\mu\nu}dx^\mu dx^\nu=-N^2(t)dt^2+a^2(t)\gamma_{ij}(x)dx^idx^j,
\end{equation}
with the spatial metric
$$\gamma_{ij}(x)dx^idx^j=\frac{dr^2}{1-kr^2}+r^2(d\theta^2+\sin^2\theta d\phi^2),$$
and generalized the reference metric from Minkowski metric to de Sitter metric,
\begin{equation}
\label{refmetric}
\eta_{ab}d\phi^a d\phi^b=-dT^2+b_k^2(T)\gamma_{ij}dX^idX^j,
\end{equation}
where the St\"{u}ckelberg fields are assumed to be $\phi^0=T=f(t)$, $\phi^i=X^i=x^i$,
so that the tensor $\Sigma_{\mu\nu}$ takes the homogeneous and isotropic form,
\begin{equation}
\label{stuckphi1}
\Sigma_{\mu\nu}={\rm Diag}\{-\dot{f}^2,\ b^2_k[f(t)]\gamma_{ij}\},
\end{equation}
and the functions $b_k(T)$ are
$$b_0(T)=e^{H_c T},\quad b_{-1}(T)=H_c^{-1}\sinh(H_cT),\quad b_1(T)=H_c^{-1}\cosh(H_cT),$$
they then found three branches of cosmological solutions, two of them correspond to the effective cosmological constant (\ref{efflambda})
and exist for spatially flat, open and closed cases. They also found a new solution \cite{tolley,langlois12}
\begin{equation}
\label{masssol1}
\frac{db_k[f]}{df}=\frac{\dot a}{N}.
\end{equation}
For the flat case, $k=0$, substituting the de Sitter function $b_0[f(t)]=e^{H_c f(t)}$ into equation (\ref{masssol1}),
we obtain the effective energy density
and pressure for the massive graviton,
\begin{gather}
\label{effrho1}
\rho_g=-m_g^2M_{pl}^2\left(1-\frac{H}{H_c}\right)\left[3(\alpha_3+4\alpha_4)\frac{H^2}{H_c^2}
-3(1+5\alpha_3+8\alpha_4)\frac{H}{H_c}+6+12\alpha_3+12\alpha_4\right],\\
\label{effp1}
\begin{split}
p_g=&m_g^2M_{pl}^2\left[-3(\alpha_3+4\alpha_4)\frac{H^3}{H^3_c}\left(1+\frac{\dot H}{H^2}\right)
+6+12\alpha_3+12\alpha_4\right.\\
&\left.-(3+9\alpha_3+12\alpha_4)\frac{H}{H_c}\left(3+\frac{\dot H}{H^2}\right)
+(1+6\alpha_3+12\alpha_4)\frac{H^2}{H^2_c}\left(3+2\frac{\dot H}{H^2}\right)\right].
\end{split}
\end{gather}
So when $H=H_c$, $\rho_g=0$.
The Friedmann equations are
\begin{gather}
\label{frweq1}
H^2+\frac{k}{a^2}=\frac{1}{3 M_{pl}^2}(\rho_m+\rho_g),\\
\label{frweq2}
2\dot H+3H^2+\frac{k}{a^2}=-\frac{1}{M_{pl}^2}(p_m+p_g).
\end{gather}
For the flat case, substituting equations (\ref{effrho1}) and (\ref{effp1}) into Friedmann equations (\ref{frweq1}) and (\ref{frweq2}),
we get \cite{tolley,langlois12}
\begin{equation}
\label{frweq4}
\begin{split}
\frac{m_g^2}{H_0^2}\frac{H(z)}{H_c}\left[-(\alpha_3+4\alpha_4)\frac{H^2(z)}{H_c^2}
+(1+6\alpha_3+12\alpha_4)\frac{H(z)}{H_c}-3(1+3\alpha_3+4\alpha_4)\right]\\
=-E^2(z)+\Omega_m(1+z)^{3(1+w_m)}-2\frac{m_g^2}{H_0^2}(1+2\alpha_3+2\alpha_4).
\end{split}
\end{equation}
\begin{equation}
\label{acceq4}
\begin{split}
\frac{\dot H}{H^2}\left\{-2E^2(z)+\frac{m_g^2}{H_c^2}E(z)\left[3(1+3\alpha_3+4\alpha_4)\frac{H_c}{H_0}-2(1+6\alpha_3+12\alpha_4)E(z)\right.\right.\\
\left.\left.+3(\alpha_3+4\alpha_4)\frac{H_0}{H_c}E^2(z)\right]\right\}=3\Omega_{m}(1+w_m)(1+z)^{3(1+w_m)},
\end{split}
\end{equation}
where $E(z)=H(z)/H_0$. The effective equation of state $w_g=p_g/\rho_g$ for the massive graviton is
\begin{equation}
\label{weff4}
w_g=-1-\frac{2E^2(z)\dot{H}/H^2+3\Omega_m(1+w)(1+z)^{3(1+w_m)}}{3[E^2(z)-\Omega_m(1+z)^{3(1+w_m)}]}.
\end{equation}

Since $\rho_g=0$ when $H(z)=H_c$, so if $H_c=H_0$, we find that $\Omega_m=1$ which
is inconsistent with current observations, therefore $H_c\neq H_0$. If $H_c<H_0$,
then we cannot recover the standard cosmology $H^2\sim \rho$ in the past unless we fine tune the value
of $m_g^2/H_0^2$ to be very small.
From equation (\ref{frweq4}), we see that the standard cosmology is recovered when $H(z)\ll H_c$ and $m_g\ll H_0<H_c$.
At very early times, $H(z)>H_c$, the universe evolves according to $H^3 \sim \rho$. If it was radiation dominated in
the very early times, then
the universe evolves faster as $a(t)\sim t^{3/4}$ instead of $t^{1/2}$.

\begin{figure}[htp]
\centerline{\includegraphics[width=4in]{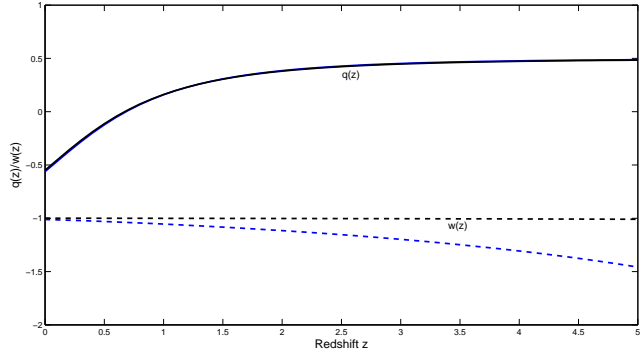}}
\caption{The evolution of the deceleration parameter $q(z)$ and the effective equation of state $w_g$ of massive graviton.
The blue lines are for the model with de Sitter metric as the reference metric and the black lines are
for the model taking $b_k(f)$ as power law form. } \label{fig1}
\end{figure}

For the special case $\alpha_3=\alpha_4=0$, Friedmann equation is simplified to
\begin{equation}
\label{frweq6}
\left(1+\frac{m_g^2}{H_c^2}\right)E^2(z)-3\frac{m_g^2}{H_cH_0}E(z)+2\frac{m_g^2}{H_0^2}=\Omega_m(1+z)^{3(1+w_m)}.
\end{equation}
At $z=0$, $E(z)=1$, we get
\begin{equation}
\label{conseq1}
\frac{m_g^2}{H_0^2}=-\frac{1-\Omega_m}{(H_0/H_c-2)(H_0/H_c-1)}.
\end{equation}
As discussed above, $H_c/H_0>1$, so $m_g^2$ must be negative for this special case. The sign of $m_g^2$ is
not important because we can always redefine the potential term so that the graviton mass is positive.
Without loss of generality, we assume that $m_g^2=-\beta_1 H_0^2$, and $H_c=\beta_2 H_0$ with $\beta_2>1$.
For the special case $\alpha_3=\alpha_4=0$, equation (\ref{conseq1}) gives
\begin{equation}
\label{conseq3}
\beta_1=\frac{(1-\Omega_m)\beta_2^2}{(2\beta_2-1)(\beta_2-1)}.
\end{equation}
In this case, we have only two free parameters $\Omega_m$ and $\beta_2$, and equation (\ref{acceq4}) gives
\begin{equation}
\label{acceq5}
\frac{\dot H}{H^2}=-\frac{3\Omega_m(1+w_m)(1+z)^{3(1+w_m)}}{2\left(1-\frac{\beta_1}{\beta_2^2}\right)E^2(z)+3\frac{\beta_1}{\beta_2}E(z)}.
\end{equation}
If $\beta_2\gg 1$, then $\beta_1=(1-\Omega_m)/2$, and the model becomes the $\Lambda$CDM model.
This is shown in Fig. \ref{fig1} for $\beta_2=20.1$ and $\Omega_m=0.3$.
Fitting this model to the three year Supernova Legacy Survey (SNLS3) sample of 472 SNe Ia data with systematic errors \cite{snls3},
and the baryon acoustic oscillation (BAO) measurements from the 6dFGS \cite{6dfgs}, the distribution of galaxies \cite{wjp} and the WiggleZ dark
energy survey \cite{wigglez}, we find the best fit values are $\Omega_m=0.27$, $\beta_2=2.69$, $\beta_1=0.71$ with $\chi^2=421.4$.
As discussed above, when $\beta_2\gg 1$, the model becomes the $\Lambda$CDM model which is independent of the value of $\beta_2$,
so $\beta_2$ cannot be constrained from above by the observational data, and we get $\beta_2\ge 1.83$ at $2\sigma$ confidence level.
The more detailed observational constraints are done in \cite{Gong:2012ny}.

\section{General cosmological solutions}

In summary, starting with the homogeneous and isotropic metric (\ref{frwmetric}) and tensor (\ref{stuckphi1}),
we obtain the Friedmann equations (\ref{frweq1}) and (\ref{frweq2})
with the effective energy density and pressure for the massive graviton,
\begin{gather}
\label{effrho2}
\rho_g=\frac{m_g^2M^2_{pl}}{a^3}(b_k[f]-a)\{6(1+2\alpha_3+2\alpha_4)a^2
-(3+15\alpha_3+24\alpha_4)a b_k[f]+3(\alpha_3+4\alpha_4)b_k[f]^2\},\\
\label{effp2}
\begin{split}
p_g=&\frac{m_g^2M^2_{pl}}{a^2}\{[6+12\alpha_3+12\alpha_4-(3+9\alpha_3+12\alpha_4){\dot f}]a^2
-2[3+9\alpha_3+12\alpha_4\\
&-(1+6\alpha_3+12\alpha_4){\dot f}]a b_k[f]
+[1+6\alpha_3+12\alpha_4-3(\alpha_3+4\alpha_4){\dot f}]b_k^2[f]\},
\end{split}
\end{gather}
and the equation of motion for the function $f(t)$ which leads to the three branches of solutions
\begin{gather}
\label{masssol4}
b_k[f(t)]=\frac{(1+6\alpha_3+12\alpha_4\pm\sqrt{1+3\alpha_3+9\alpha_3^2-12\alpha_4})}{3(\alpha_3+4\alpha_4)}a(t),\\
\label{masssol5}
\frac{db_k[f]}{df}=\frac{\dot a}{N}.
\end{gather}
When we take the solution (\ref{masssol4}), we get the effective cosmological constant solution (\ref{efflambda})
independent of the choice of spatial curvature $k$. For $k=1$, the solution was obtained in \cite{gumruk11} by taking
the reference metric $\eta_{ab}$ as Minkowski and the same homogeneous and isotropic tensor $\Sigma_{\mu\nu}$ (\ref{stuckphi1}) with $b_k[f(t)]=f(t)$. The same solution (\ref{efflambda}) was obtained in \cite{kobayashi12} for all values of $k$
for the particular parameters (\ref{1parm}), but the tensor $\Sigma_{\mu\nu}$ is not homogeneous and isotropic.
For the flat case $k=0$, Gratia, Hu and Wyman obtained the same cosmological constant solution (\ref{efflambda}) by using
another inhomogeneous and anisotropic tensor $\Sigma_{\mu\nu}$ \cite{gratia12}. Motohashi and Suyama obtained the cosmological
constant solution for the $k=0$ case with isotropic forms for both the physical and reference metrics \cite{Motohashi:2012jd}.
However, the cosmological constant solution (\ref{efflambda}) is just the consequence of the equation of motion of
the function $f(t)$ once we assumed the homogeneous and isotropic form for the metric (\ref{frwmetric}) and the tensor $\Sigma_{\mu\nu}$ (\ref{stuckphi1}).
Since the cosmological constant solution (\ref{efflambda}) was obtained by different methods for different sepcial cases, this suggests that
this solution exists for the general case. The method proposed in \cite{tolley,langlois12} not only gives the solution for the general case,
but also gives additional new dynamic solutions.
This suggests that the solution (\ref{masssol5}) should be quite general even without the assumption of the reference metric $\eta_{ab}$ as de Sitter.
Therefore, we propose that more solutions can be found with equations (\ref{frwmetric}) and (\ref{stuckphi1})
by assuming more general form of $b_k[f(t)]$. Note that the specific form of $\Sigma_{\mu\nu}$ in equation (\ref{stuckphi1})
may be obtained from Minkowski, de Sitter, or isotropic reference metrics.

Follow the above argument, we assume a power law form $b_k[f(t)]=(H_c f(t))^{\gamma/(\gamma-1)}$ with
$\gamma>1$,
then the solution to equation (\ref{masssol5}) is
\begin{equation}
\label{masssoln}
b_k[f(t)]=\left(\frac{a(\gamma-1)H}{\gamma H_c}\right)^\gamma,
\end{equation}
and the effective energy density becomes
\begin{equation}
\label{rhogn}
\begin{split}
\rho_g=&3m_g^2M_{pl}^2\left[3(1+3\alpha_3+4\alpha_4)\left(\frac{\gamma-1}{\gamma}\right)^\gamma a^{\gamma-1}\left(\frac{H}{H_c}\right)^\gamma\right.\\
&-(1+6\alpha_3+12\alpha_4)\left(\frac{\gamma-1}{\gamma}\right)^{2\gamma} a^{2\gamma-2}\left(\frac{H}{H_c}\right)^{2\gamma}\\
&\left.-2(1+2\alpha_3+2\alpha_4)+(\alpha_3+4\alpha_4)\left(\frac{\gamma-1}{\gamma}\right)^{3\gamma} a^{3\gamma-3}\left(\frac{H}{H_c}\right)^{3\gamma}\right].
\end{split}
\end{equation}
The effective pressure of massive graviton is
\begin{equation}
\label{pgn}
\begin{split}
p_g=&m_g^2M_{pl}^2\left\{6(1+2\alpha_3+2\alpha_4)-3(1+3\alpha_3+4\alpha_4)\left(\frac{\gamma-1}{\gamma}\right)^\gamma a^{\gamma-1}\left(\frac{H}{H_c}\right)^\gamma
\left[2+\gamma\left(1+\frac{\dot H}{H^2}\right)\right]\right.\\
&+(1+6\alpha_3+12\alpha_4)\left(\frac{\gamma-1}{\gamma}\right)^{2\gamma} a^{2\gamma-2}\left(\frac{H}{H_c}\right)^{2\gamma}
\left[1+2\gamma\left(1+\frac{\dot H}{H^2}\right)\right]\\
&\left.-3\gamma(\alpha_3+4\alpha_4)\left(\frac{\gamma-1}{\gamma}\right)^{3\gamma} a^{3\gamma-3}\left(\frac{H}{H_c}\right)^{3\gamma}
\left(1+\frac{\dot H}{H^2}\right)\right\}.
\end{split}
\end{equation}
Again when the hubble parameter is in the range $H_0<H(z)<H_c$, the standard cosmology is recovered. To have a long history
of matter and radiation domination, we require $H_0\ll H_c$.

For the special case $\alpha_3=\alpha_4=0$, Friedmann equations are
\begin{equation}
\label{frweqn1}
E^2(z)-\frac{\beta_1}{\beta_2^{2\gamma}}\left(\frac{\gamma-1}{\gamma}\right)^{2\gamma} a^{2\gamma-2}E^{2\gamma}(z)-2\beta_1+
3\frac{\beta_1}{\beta_2^{\gamma}}\left(\frac{\gamma-1}{\gamma}\right)^\gamma a^{\gamma-1}E^\gamma(z)=\Omega_m(1+z)^{3(1+w_m)},
\end{equation}

\begin{equation}
\label{acceqn1}
\frac{\dot H}{H}=-1+\frac{\frac{\beta_1}{\beta_2^{2\gamma}}\left(\frac{\gamma-1}{\gamma}\right)^{2\gamma}a^{2\gamma-2}E^{2\gamma}
-6\frac{\beta_1}{\beta_2^{\gamma}}\left(\frac{\gamma-1}{\gamma}\right)^{\gamma}a^{\gamma-1}E^{\gamma}-E^2+6\beta_1}{
-2\gamma\frac{\beta_1}{\beta_2^{2\gamma}}\left(\frac{\gamma-1}{\gamma}\right)^{2\gamma}a^{2\gamma-2}E^{2\gamma}+3\gamma
\frac{\beta_1}{\beta_2^{\gamma}}\left(\frac{\gamma-1}{\gamma}\right)^{\gamma}a^{\gamma-1}E^{\gamma}+2E^2},
\end{equation}
with
\begin{equation}
\label{constreqn1}
\beta_1=\frac{(1-\Omega_m)\beta_2^{2\gamma}}{\left[\left(\frac{\gamma-1}{\gamma}\right)^\gamma -\beta_2^\gamma\right]\left[\left(\frac{\gamma-1}{\gamma}\right)^\gamma -2\beta_2^\gamma\right]}.
\end{equation}
In this case, we have three free parameters $\Omega_m$, $\gamma$ and $\beta_2$. Again if $\beta_2\gg 1$,
$\beta_1=(1-\Omega_m)/2$, the model is equivalent to the $\Lambda$CDM model and is independent of the values of $\beta_2$ and $\gamma$.
This is shown in Fig. \ref{fig1} for $\beta_2=20.1$ and $\Omega_m=0.3$.

\section{Conclusions}

In conclusion, more general cosmological solutions which are consistent with
the observational data can be found by taking homogeneous and isotropic form
for both the metric $g_{\mu\nu}$ and the tensor $\Sigma_{\mu\nu}$ without specifying the form of reference metric,
even though the tensor $\Sigma_{\mu\nu}$ may be obtained with Minkowski, de Sitter or isotropic reference metrics.
In addition to the cosmological constant solution, more richer dynamics can be found in these solutions.
The mass of graviton is in the order of $((1-\Omega_m)/2)^{1/2}H_0$.

\begin{acknowledgments}

This work was partially supported by
the National Basic Science Program (Project 973) of China under
grant No. 2010CB833004, the National Nature Science Foundation of China under grant Nos. 10935013 and 11175270,
and the Fundamental Research Funds for the Central Universities.

\end{acknowledgments}


\end{document}